\documentclass[twocolumn,english]{revtex4}
\usepackage[T1]{fontenc}
\usepackage[latin9]{inputenc}
\usepackage{graphicx}

\makeatletter

\usepackage{babel}
\makeatother

\begin{document}

\title{Role of the self-interaction error in studying first principles
 chemisorption on graphene}

\author{Simone Casolo$^{a}$}

\email{simone.casolo@unimi.it}

\author{Espen Flage-Larsen$^b$, 
Ole Martin Løvvik$^{b,c}$,
George R. Darling$^d$,
Gian Franco Tantardini$^{a,e}$}

\affiliation{
$^a$Dipartimento di Chimica Fisica ed Elettrochimica, Università degli
Studi di Milano, via Golgi 19, 20133 Milan, Italy.\\
$^b$Department of Physics, University of Oslo, P.O. Box 1048 Blindern, 
NO-0316 Oslo, Norway\\
$^c$SINTEF, Materials and Chemistry, Forskningsvn. NO-0314 Oslo, Norway\\
$^{d}$Surface Science Research Centre, Department of Chemistry, 
The University of Liverpool, Liverpool L69 7ZD, United Kingdom\\
$^e$CIMAINA, Interdisciplinary Center of Nanostructured Materials and 
Interfaces, University of Milan
}

\begin{abstract}

Adsorption of gaseous species, and in particular of hydrogen atoms, 
on graphene is an important process for the chemistry of this material.
At the equilibrium geometry, the H atom is covalently bonded to a 
carbon that puckers out from the surface plane. 
Nevertheless the \emph{flat} graphene geometry 
becomes important when considering the full sticking dynamics. 
Here we show how GGA-DFT predicts a wrong spin state for this geometry, 
namely $S_z$=0 for a single H atom on graphene. 
We show how this is caused by 
the self-interaction error since the system shows fractional electron 
occupations in the two bands closest to the Fermi energy. 
It is demonstrated how the use of hybrid functionals or the 
GGA+$U$ method an be used to retrieve the correct spin solution although 
the latter gives an incorrect potential energy curve. 

\end{abstract}
\maketitle

\section{Introduction}

Thanks to its peculiar semi-metallic band structure, graphene
is a very promising material for future silicon-free nano electronics
\cite{castroneto08,Avouris07}. 
In particular, graphene's charge-carriers mobility is 
indeed extraordinarily high, with mean-free paths in the 
order of microns\cite{Bolotin08,Schedin07}.
However, for the fabrication of
logic devices the absence of a band gap is a major 
issue since it does not allow a complete turn-off of the current, hence 
high on-off ratios\cite{Avouris07,Novoselov07}.\\
One straightforward  possibility for graphene band-gap engineering 
is to use the adsorption of radicals or small molecules to create 
$\pi$ defects, breaking the equivalence of its two sublattices.
%Otherwise, one can localize the wave function by cutting out a 
%nano-ribbon,
%a process now possible via chemical etching by radical species. 
%Moreover radicals adsorption on graphene is also a key step 
%in the chemical etching for ``cutting'' out graphene nano-ribbons: 
%another way to induce a gap by electron localization\citep{Han07}.\\
In this perspective several studies have been performed on 
isolated\cite{Jeloaica1999,Sha&Jackson2002}, 
clusters\cite{casolo09,Hornekaer2006,Hornekaer2006a,Boukhvalov2008} 
and even superlattices of adsorbates bonded on graphene, 
and most of them are based on the density functional theory (DFT).
Among these hydrogen atoms adsorption is by far the most studied
case, also because of the many implications in other 
fields\cite{Gould63,Hollenbach70}.\\
Hydrogen chemisorption dynamics on graphene is not yet completely 
understood. In the adiabatic picture when a H atom impinges on graphene
and binds at its top site it induces one carbon atom to move out from the 
plane, ``puckering'' the surface.
On the other hand, if the  
incoming species moves fast enough toward the graphene layer,  
sticking can occur faster than surface reconstruction; in this case the 
substrate can be considered as rigid. The planar 
geometry may thus play an importance role in the adsorption dynamics, 
and the accuracy of DFT calculations
in this case is crucial in order to build reliable potential 
energy surfaces for quantum dynamics studies. \\
In this article we analyze the spin properties of the substrate when 
a H atom is forced to chemisorb on a \emph{flat} graphene sheet.
We show how the loss of spin polarization is a fictitious feature of 
semi-local generalized gradient approximation (GGA) 
functionals due to the self-interaction error (SIE). 
Then we will show how the correct ground state magnetization 
can be achieved using hybrid functional or GGA+$U$ approaches, even 
though the latter approach fails in reproducing the correct H-graphene
potential energy curve. 

\section{Computational Methods}

In brief, periodic density functional theory as implemented in 
the VASP package \citep{VASP1,VASP2} has
been used throughout. 
A GGA-PBE functional was used and the 
plane wave basis set was limited to a 500 eV energy cutoff. 
For the inner electrons we rely on the frozen core approximation 
using PAW pseudo-potentials \citep{PAW1}.\\
The reciprocal space was sampled by $\Gamma$ centered k-point grids,
whose meshes were chosen depending on the supercell size, in any case 
never more sparse than 6x6x1. 
The graphene unit supercells used here range from a 2x2
to a 5x5: all of them have a vacuum region along the \emph{c} axis of 20
\AA~ in order to guarantee a vanishing interaction between periodically
repeated images.\\ It has been shown
recently that for hydrogen adsorption a 5x5 supercell is still not
big enough to extract adsorption energies at meV accuracy 
\citep{casolo09}, although this goes beyond the aim of this work.
Further details about the computational setup can be found in 
ref.\citep{casolo09}.

\section{Results and Discussion}

When a radical species chemisorbs onto graphene or graphite surfaces 
the most favorable outcome is the formation of 
a covalent bond with one of the surface carbons, \emph{i.e.} 
at a top site.
The simplest case of radical is a single (neutral) hydrogen atom.
Applying Valence Bond (VB) arguments to this reaction 
one can predict that as
the atom approaches the substrate plane it interacts with a $\pi$ 
electron of graphene, triggering
orbital re-hybridization of the C atom from a planar $sp^{2}$
to a partially tetrahedral $sp^{3}$ 
configuration\cite{Sha&Jackson2002,Jeloaica1999}.
At long range the adsorbate - substrate interaction is 
purely repulsive since there is no unpaired electron on graphene 
available to bind hydrogen. 
At short range however a low-lying spin-excited state in which 
two $\pi$ electrons lying on opposite,
non-overlapping ends of a benzene ring would give rise
to an attractive, barrierless interaction with the H 1$s$ lone 
electron. Hence, 
an avoided crossing between these two doublet curves occurs
giving rise to an activation barrier to chemisorption (see Fig.5 
in ref.\cite{casolo09}).
When the graphene sheet is allowed to reconstruct
this process reads as a $sp^2-sp^3$ orbital re-hybridization 
of the carbon involved in the bonding that turns
partially tetrahedral, puckering out 0.6 
\AA~from the layer plane.\\
The graphene lattice is a bipartite system, made of two equivalent 
sublattices, each made of every second carbon atom. 
The equivalence of the two honeycomb sublattices is responsible 
for the particle-hole symmetry in graphene, and for the peculiar 
conical intersection at $E_F$ between the valence and conduction bands 
\cite{Weiss58}.
A chemisorbed species creates a defect in the aromatic network, hence 
an imbalance between the number of occupied sites of the two $\pi$ 
sublattices ($n_A$ and $n_B$ respectively).
According to a theorem formulated by Inui \emph{et. al.} 
within tight-binding theory,
whenever an imbalance (vacancy) is introduced in a bipartite 
lattice this gives rise to $|n_A-n_B|$ zero-energy midgap states, 
localized on one sublattice only \cite{Inui1994}.
Moreover, following the second
Lieb's theorem \cite{lieb}, since the total number of electrons is
odd the total magnetization for non-metallic systems has
to be S$^2=|n_A-n_B|/2$.  
Thus, for a single defect S$_z$=1/2, or 1 $\mu_{B}$.\\
DFT calculations confirm this picture: the hydrogen atom introduces 
a defect in one of the two sublattices, breaking one among the many
 aromatic bonds around the tetrahedral carbon. 
This implies that an unpaired electron can be delocalized by a 
``bond switching''
process along the other sublattice, made of every second carbon atom. 
In the energy spectrum this reads as a flat band at the Fermi
level \emph{i.e.} in a midgap state occupied by one
single spin projection only\cite{Wehling08,casolo09}.\\ 
According to previous studies \cite{Sha&Jackson2002,Jeloaica1999}
we found at the GGA-DFT level 
that keeping the substrate planar thwarts the $sp^{2}-sp^{3}$
re-hybridization; this is enough to weaken the attractive C-H interaction,
but also to corrupt the system's aromatic character. 
Nevertheless, because the VB arguments concerning the crossing 
of two spin states hold, only a meta-stable C-H bond can form.\\
From our DFT calculations we notice how the total spin for H adsorbed 
on flat graphene is lower than expected at the local minimum geometry. 
%Indeed, following the adsorption path 
%along the surface normal direction it is possible to see how as soon 
%as a crictical height is reached, the total spin of the system drops. 
In Fig.\ref{fig:Total-spin-vs} is shown the value of magnetization
(left panel) 
along the adsorption path for several surface coverages, together with
the total energies for the spin-polarized and unpolarized solutions 
(right panel). 
When the radical is far from the surface
its magnetic moment is correctly 1 $\mu_{B}$: this corresponds
to an electron lying in the H 1$s$ orbital, while graphene 
electronic structure 
remains intact. As the atom approaches
the graphene layer along the normal direction, 
at a given critical height  from the surface ($z_{c}(H)\simeq1.25$ \AA)
the system's spin drops. The minimum value for the 
total spin depends upon the coverage. When coverage is low enough 
the spin is eventually quenched down to zero. 
Pushing the adsorbate closer to the carbon atom the magnetization tends
to increase again. 
We tested that the same picture also holds for other small organic 
radicals with slightly different critical heights, weakly
dependent upon the supercell size (coverage).\\
\begin{figure}
\begin{centering}
\includegraphics[clip,width=1\columnwidth]{fig1}
\par\end{centering}

\caption{Left panel: total spin vs.
 distance from the surface for a H atom 
Adsorbing on flat graphene. The coverages are the following: 
$\Theta$=0.031 (full line), 0.055 (dashed) and 0.125 ML (dotted).\\
Right panel: adsorption potential curves for H (full line)  
together with the same curves for the spin unpolarized case
(squares) and for the fixed magnetization $1\ \mu_{B}$ (circles).
The full line shows the adiabatic (non-constrained) curve.
Zero energy here is set asymptotically away from graphene. 
\label{fig:Total-spin-vs}}

\end{figure}
\begin{figure}
\begin{centering}
\includegraphics[clip,width=0.95\columnwidth]{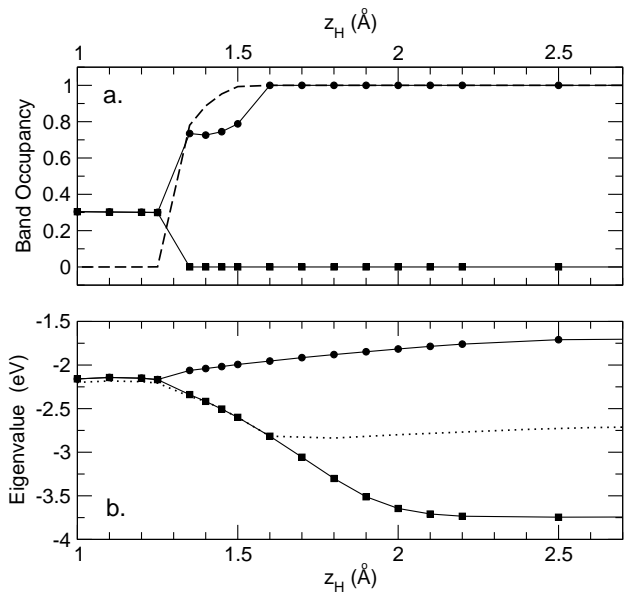}
\par\end{centering}

\caption{\emph{a}: Occupancy for the spin up (circles) 
and spin down (squares) bands closest to $E_F$ along
the whole adsorption path. 
Dashed line: total magnetization (in $\mu_{B}$).
\emph{b}: Eigenvalues for the bands shown above, the dotted
 line represents the value of the Fermi energy. 
\label{fig:Adsorption-of-atomic}}

\end{figure}
When comparing the adsorption curves computed with a magnetization
fixed to 0 and 1 $\mu_{B}$ as in Fig.\ref{fig:Total-spin-vs} (right panel), it
is clear that the non-magnetic (non-polarized) solution becomes more
stable than the magnetic one beyond the critical point $z_c$.\\
%Looking at the density of states as the atom gets closer
%to the surface its state starts to hybridize with the $\pi$ cloud,
%in particular with components of the spin minority, and a band-gap
%opens. The result is the formation of a state that is symmetric respect
%to both spin components, instead to have a ``defect'' state 
%of a single spin component in the gap. If one forces the system to keep
%a doublet multiplicity all over the adsorption process then spin majority
%and minority hybridize differently and the defect state occupied by
%the radical electron only is formed.\\
To a closer look (see Fig.\ref{fig:Adsorption-of-atomic}b) 
we notice that the occupied hydrogen $s$ orbital
and its (empty) affinity level get closer in energy when 
approaching graphene consistently with the Newns-Anderson model 
\cite{NewnsAnderson2,NewnsAnderson1}.
Then around $z_{c}$ they approach $E_{F}$, become degenerate
and equally occupied by a fractional number of electrons. 
In the system's density of states,
a gap opens at the point of the spin quenching together with the 
formation of a broad partially occupied peak at the Fermi energy,
 symmetric for both spin projections. \\
If the host were a metal, then spin-flip scattering between conduction
electrons and a magnetic impurity might lead to a singlet ground
state (Kondo effect). 
However in our case there is not a delocalized free-electron-gas-like 
surface state (a Shockley state) that is free to screen the impurity. 
Moreover the $s$ electron does not belong to a localized 
orbital decoupled from the  substrate such as for $d$ metals.
The Kondo effect is a many body feature that is not well described by 
DFT. Anyway this technique can describe correctly the spin quenching on 
metallic substrates, such as for a hydrogen atom
impinging on Cu, Ag or Al(111) surfaces where this
spin transition is a signature of non-adiabatic effects 
\cite{Lindenblatt2006,Mizielinski2008}.\\
%On metals, once H approaches the surface its 1\emph{s}
%spin orbital hybridizes with the metal's Bloch states, shifts in energy
%towards the Fermi level, and broadens. This is what we see in DFT
%calculations of H adsorption on a 10 layers slab representing the
%Cu(100) surface, in which the hydrogen electron does not split in
%two bands as shown in fig.\ref{fig:Adsorption-of-atomic}\emph{b}
%but rather goes to occupy many of the metallic bands closer to the
%Fermi level. At the transition point, the electron merges totally
%with the metal states that act as a reservoir\citep{Lindenblatt2006},
%hence the local electronic spin moment of the hydrogen atom goes to
%zero. In case of a graphene there are only electronic states localized
%on each of the two sublattices and the density of states for the $\pi$
%bands vanishes at the Fermi level, so any hybridization with H 1\emph{s}
%orbitals will be less efficient having no reservoir at the Fermi energy.
We found that radical adsorption on planar graphene
is a situation in which the (many-electron) 
self-interaction error (thereafter SIE, 
also known as delocalization error\cite{CohenScience}) 
is particularly severe. 
Following the notation in ref\cite{Gritsenko2000}
this is a chemical reaction with two centers ($m=2$) and one electron 
($n=1$),
since none of the electrons of graphene may be directly involved in
bonding if they cannot re-hybridize to a tetrahedral \emph{sp$^{3}$}
state. Systems with a fractional $n/m$ ratio are also known to exhibit
of large static electronic correlation \cite{ZhangYang98}.\\ 
A well known system with $n/m=1/2$ 
is the H$_{2}^{+}$ molecular ion dissociation.
When the two atoms are far apart both 
the local density approximation (LDA) and GGA local functionals 
give as the most stable ground state half of an electron on each of the 
two degenerate atomic orbitals, which is physically
not correct.
Asymptotically the orbitals on the two fragments are degenerate, so this 
fractional occupation solution should be degenerate with any other
possible electronic arrangement such as one filled and one empty orbital 
\cite{MoriSanchez09}. 
%The H$_{2}^{+}$ molecule is a typical example of open system 
%with a fluctuating electron number.
Fractional charge (and spin) on the two fragments is due to the 
over-delocalization that is a direct consequences of the SIE. 
Indeed the LDA and GGAs functionals are designed to correctly reproduce 
the system's total density and the on-top pair density, 
but fail in pathological systems to reproduce spin densities 
because of the SIE \cite{Burke98,Perdew95}.
This non-integer orbital occupation is directly related 
to the break down of sum rules over the exchange hole density 
${n_x}({\bf{r,r'}})$ 
\begin{equation}
\int{n_x}({\bf{r,r'}})d{\bf{r'}}=
\sum_{\alpha,\sigma}f_{\alpha,\sigma}
\frac{n_{\alpha,\sigma}({\bf{r}})}{n({\bf{r}})}
\end{equation}
that when the orbital occupation $f$ becomes fractional sums to a 
value within larger than the correct -1.
Thus local functionals give an energy whose behaviour is convex 
for a fractional number of electrons instead of being linear 
according to Janak theorem \cite{Perdew1982}. For this reason 
for open systems a delocalized situation 
with fractional charge turns out to be more favored \cite{CohenScience}.\\
For H adsorbed on flat graphene the spin quenching is similarly due
to a fractional spin situation: the ``splitting'' of one electron
in two different \emph{degenerate} 
bands (originating from 
H \emph{s} and C $p_z$), with opposite spin projections
(see Fig.\ref{fig:Adsorption-of-atomic}).
Here the occupation number in these bands can fluctuate, a sign that
SIE is particularly severe \cite{Perdew07}.
This picture is confirmed by the convex behaviour of the system energy
for fractional band occupation as shown in Fig.\ref{fig:frac}, obtained 
by constraining a given occupation within the two bands. 
Note that two degenerate orbitals every arrangement of electrons 
(even for fractional numbers) should be perfectly 
degenerate. In this case however, the energy minimum lays exactly 
at the unpolarized solution: 0.5 occupancy of the two bands.\\
\begin{figure}
\begin{centering}
\includegraphics[clip,width=0.99\columnwidth]{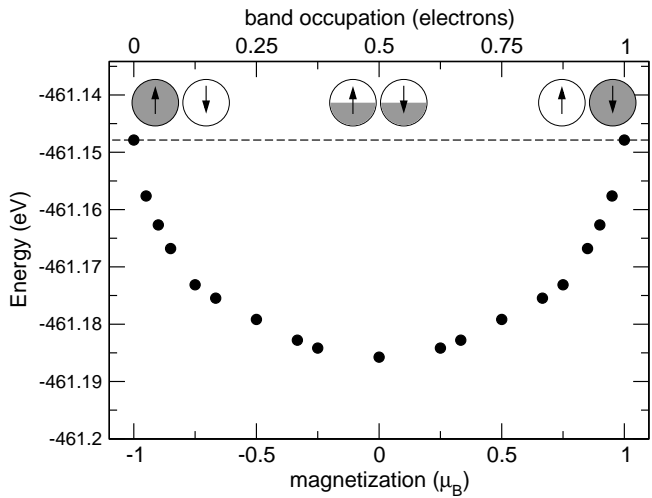}
\par\end{centering}

\caption{Calculated total energy vs. magnetization for one
 H atom adsorbed on flat graphene. 
When the magnetization is -1 or 1 $\mu_{B}$ one of the two degenerate 
bands is occupied while the other is empty. For the non-polarized, 0 $\mu_{B}$,
case both bands are occupied by half of an electron each. 
The full line represents 
the correct degenerate behaviour for fractional electron numbers; 
filled circles are the GGA-DFT results.}
\label{fig:frac}

\end{figure}
Another indication of the SIE is the worsening of the spin quenching at
low coverages, hence for larger and larger super-cells, as shown before 
in Fig.\ref{fig:Total-spin-vs}. 
With the SIE being cause of the delocalization, the 
fractional occupancy might not be maximal when the unit cell is not 
large enough 
to accommodate all the delocalized electron density \cite{MoriSanchez08}.\\
A major difference with the H$_{2}^{+}$ prototype case is that here
 there is no fractional spin asymptotically for H
 since the involved orbitals here
lay far below graphene Fermi energy. For other and more electronegative
monovalent species such as F and OH fractional charges appear
also asymptotically, similarly to the case of dissociations
of heteronuclear diatomics \cite{Ruzsinszky06}.\\ 
\begin{figure}
\begin{centering}
\includegraphics[clip,width=0.98\columnwidth]{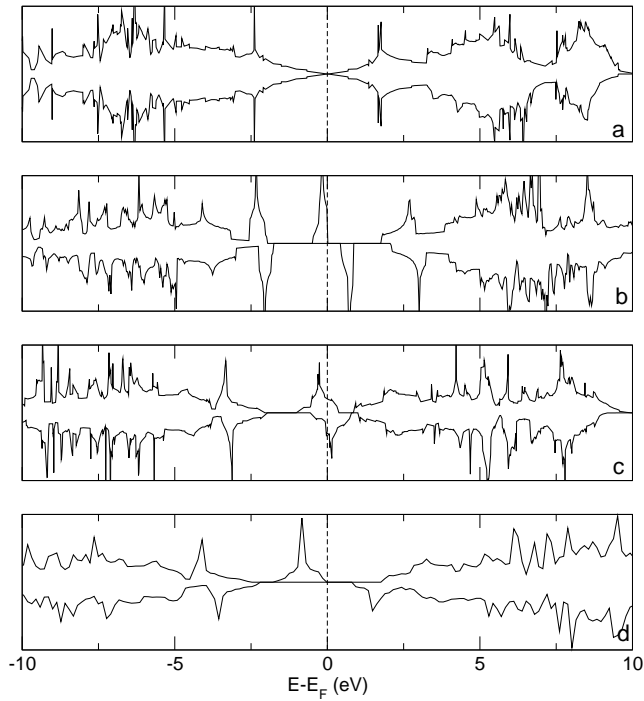}
\par\end{centering}

\caption{Density of states for the H-flat graphene system 
obtained for a 2x2 supercell.
PBE results for: a) clean graphene, b) H adsorbed on reconstructed 
(puckered) graphene, c) H adsorbed on flat graphene (M=0$\mu_{B}$). 
Hybrid-PBE0 results ($1\mu_{B}$) are shown in d) for comparison.
As a guide to the eye the Fermi energy is shown as dashed vertical line.}
\label{fig:DOS}

\end{figure}
To prove further that the failure in representing the total spin 
of the system comes
from the approximate nature of the density functionals we tested
the performances of a (non-local) hybrid functional. 
Hybrid functionals combine
the GGA exchange and correlation term (convex for fractional electron
number) with the Hartree-Fock (HF) ``exact'' exchange term. 
Since HF energies have instead a concave behaviour for fractional electron
numbers they often yield over-corrected results \cite{MoriSanchez08}.
For this reason the use of a fraction (usually one fourth)
of exact exchange to correct the DFT E$_{xc}$ functional
gives much better results, 
at least for non-metallic systems \cite{Marsman08}.
We then chose the PBE0 functional, which mixes 25\% of HF 
with 75\% of PBE exchange, and uses the full PBE correlation\cite{PBE0}:
 \[
E_{xc}^{PBE0}=E_{xc}^{PBE}+0.25(E_{x}^{HF}-E_{x}^{PBE})
\]
%in a way to compare this with the of the local GGA-PBE.
Hybrid functionals are orbital dependent, \emph{i.e.} non-local in space: 
this is a major issue
when employing plane wave codes where the number of orbitals 
depends upon the supercell volume. The computational effort needed for
this kind of calculations is thus much larger compared to GGA. Hence
we could not reach low coverages, restricting our system to 
a 2x2 supercell ($\Theta$=0.125 ML).\\
Within hybrid functional DFT the system's total spin for H 
% 1 \AA~from the 
on flat graphene layer is 1 $\mu_{B}$ ($S_z$=1/2), 
and the associated spin density is correctly localized on one 
sublattice only.
A comparison between the density of states computed with PBE and PBE0 
is shown in Fig.\ref{fig:DOS}. 
GGA can represent pretty well the adiabatic 
chemisorption mechanism, 
namely the band gap opening and the zero energy midgap state
that splits by the exchange interaction into a filled and an empty 
states with opposite spin (Fig\ref{fig:DOS}\emph{a} 
and \emph{b}).
For the flat geometry, GGA gives a broad feature straddling across the 
Fermi energy (Fig\ref{fig:DOS}\emph{c}). 
For larger supercells (at lower coverages) the peak 
becomes fully symmetric for the two spin projections, giving rise to the
un-polarized state. 
On the other hand PBE0 can reproduce well the occupied midgap state.  
As it is widely known standard DFT
tend to underestimates the band gaps, again due to the 
SIE\cite{CohenBandGap}: for this
reason the PBE0 band-gap is about 50\% larger than that of PBE.\\ 
It has also been proposed that an ``on site'' 
repulsion term such as in the LSDA+$U$ \cite{LSDAU} 
approach can help to control the SIE in case of 
fractional occupation \cite{Cococcioni05}.
The on site Coulomb term $U$ acts as a penalty 
for the occupation of the two degenerate bands at the Fermi energy,
and can thus reproduce the correct ground state. Note that
LSDA+$U$ and GGA(PBE)+$U$, as implemented here, 
give practically identical results in this case.
This approach is much less computationally expensive than the hybrid 
functionals, so we could study the full adsorption path. As for PBE0
the total magnetization is 1 $\mu_{B}$ at every C-H distance, and the 
spin density correctly is localized either on the H atom 
or on one graphene sublattice.
From our tests a Coulomb interaction of 15 eV was enough
to retrieve the correct ground state spin: a value not far from 20.08
eV, already successfully used to describe carbon $\pi$ electrons
in similar approaches \cite{Pedersen04,Pedersen08}.\\
\begin{figure}
\begin{centering}
\includegraphics[clip,width=0.8\columnwidth]{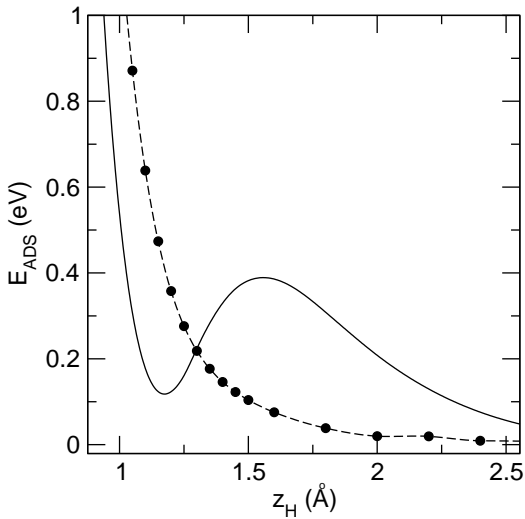}
\par\end{centering}
\caption{Comparison between GGA (full line) and GGA+$U$ (dashed line) 
potential energy curves along the adsorption path. Note how GGA+$U$ do 
not show the correct activation barrier arising from the avoided crossing 
of the two curves shown in the right panel of Fig.\ref{fig:Total-spin-vs}}
\label{fig:gga+u}
\end{figure}
We would like to stress that 
the GGA+$U$ approach is not a rigorous way to avoid self interaction,
 and hence
some care has to be taken when interpreting these results. While it
is relatively easy to predict which has to be the correct total spin
of the system upon physical arguments, it is more challenging to judge 
the quality of total energies.
%It is known that self interaction,
%as well as static correlation, tend to underestimate the barrier height
%\cite{Csonka1998}. \\
Indeed, it can be seen from Fig.\ref{fig:gga+u} that the effect of the on-site
\emph{U} term is to lower the activation 
barrier at $z_{c}$ making the adsorption
curve to a fully repulsive interaction.  
Being the chemisorption potential energy curve for H on graphene the 
result of the interplay of two an attractive and a repulsive 
doublet spin states their avoided crossing has necessarily
to occur. For this reason the GGA+$U$ results, despite showing
the correct magnetization, cannot
represent the correct H-graphene interaction in all its aspects.

\section{Conclusions}

The chemisorption of a hydrogen atom onto a flat graphene sheet 
has been studied within semi-local GGA-DFT.
Contrary to the adiabatic case the substrate $sp^{2}-sp^{3}$
re-hybridization is limited, so adsorbate and substrate bands 
become degenerate at a given critical C-H distance, $z_{c}$, where 
the total magnetization drops to zero.\\
This spin transition is due to a fractional spin configuration:
a fictitious effect induced by the self-interaction error. 
To overcome this issue it is possible to use a hybrid functional such as 
PBE0. The GGA+$U$ approach can reproduce the correct 
magnetization, but leads to qualitatively incorrect potential energy 
curves.

\section{Acknowledgments}

The authors thank the NOTUR consortium 
for providing computational resources .
S. C. acknowledges the University of Oslo for the
hospitality during his stay and Rocco Martinazzo and George Darling
for the many useful comments.

%Here a small summary and conclusion.

\bibliographystyle{apsrev}

\begin{thebibliography}{41}
\expandafter\ifx\csname natexlab\endcsname\relax\def\natexlab#1{#1}\fi
\expandafter\ifx\csname bibnamefont\endcsname\relax
  \def\bibnamefont#1{#1}\fi
\expandafter\ifx\csname bibfnamefont\endcsname\relax
  \def\bibfnamefont#1{#1}\fi
\expandafter\ifx\csname citenamefont\endcsname\relax
  \def\citenamefont#1{#1}\fi
\expandafter\ifx\csname url\endcsname\relax
  \def\url#1{\texttt{#1}}\fi
\expandafter\ifx\csname urlprefix\endcsname\relax\def\urlprefix{URL }\fi
\providecommand{\bibinfo}[2]{#2}
\providecommand{\eprint}[2][]{\url{#2}}

\bibitem[{\citenamefont{{Castro Neto} et~al.}(2009)\citenamefont{{Castro Neto},
  Guinea, Peres, Novoselov, and Geim}}]{castroneto08}
\bibinfo{author}{\bibfnamefont{A.~H.} \bibnamefont{{Castro Neto}}},
  \bibinfo{author}{\bibfnamefont{F.}~\bibnamefont{Guinea}},
  \bibinfo{author}{\bibfnamefont{N.~M.~R.} \bibnamefont{Peres}},
  \bibinfo{author}{\bibfnamefont{K.~S.} \bibnamefont{Novoselov}},
  \bibnamefont{and} \bibinfo{author}{\bibfnamefont{A.~K.} \bibnamefont{Geim}},
  \bibinfo{journal}{Rev. Mod. Phys.} \textbf{\bibinfo{volume}{81}},
  \bibinfo{pages}{109} (\bibinfo{year}{2009}).

\bibitem[{\citenamefont{Avouris et~al.}(2007)\citenamefont{Avouris, Chen, and
  Perebeinos}}]{Avouris07}
\bibinfo{author}{\bibfnamefont{P.}~\bibnamefont{Avouris}},
  \bibinfo{author}{\bibfnamefont{Z.}~\bibnamefont{Chen}}, \bibnamefont{and}
  \bibinfo{author}{\bibfnamefont{V.}~\bibnamefont{Perebeinos}},
  \bibinfo{journal}{Nat. Nanotech.} \textbf{\bibinfo{volume}{2}},
  \bibinfo{pages}{605} (\bibinfo{year}{2007}).

\bibitem[{\citenamefont{Bolotin et~al.}(2008)\citenamefont{Bolotin, Sikes,
  Jang, Klima, Fudenberg, Hone, Kim, and Stromer}}]{Bolotin08}
\bibinfo{author}{\bibfnamefont{K.}~\bibnamefont{Bolotin}},
  \bibinfo{author}{\bibfnamefont{K.}~\bibnamefont{Sikes}},
  \bibinfo{author}{\bibfnamefont{Z.}~\bibnamefont{Jang}},
  \bibinfo{author}{\bibfnamefont{M.}~\bibnamefont{Klima}},
  \bibinfo{author}{\bibfnamefont{G.}~\bibnamefont{Fudenberg}},
  \bibinfo{author}{\bibfnamefont{J.}~\bibnamefont{Hone}},
  \bibinfo{author}{\bibfnamefont{P.}~\bibnamefont{Kim}}, \bibnamefont{and}
  \bibinfo{author}{\bibfnamefont{H.~L.} \bibnamefont{Stromer}},
  \bibinfo{journal}{Solid State Commun.} \textbf{\bibinfo{volume}{143}},
  \bibinfo{pages}{351} (\bibinfo{year}{2008}).

\bibitem[{\citenamefont{Schedin et~al.}(2007)\citenamefont{Schedin, Geim,
  Morozov, Hil, Blake, Katsnelson, and Novoselov}}]{Schedin07}
\bibinfo{author}{\bibfnamefont{F.}~\bibnamefont{Schedin}},
  \bibinfo{author}{\bibfnamefont{A.~K.} \bibnamefont{Geim}},
  \bibinfo{author}{\bibfnamefont{S.~V.} \bibnamefont{Morozov}},
  \bibinfo{author}{\bibfnamefont{E.~W.} \bibnamefont{Hil}},
  \bibinfo{author}{\bibfnamefont{P.}~\bibnamefont{Blake}},
  \bibinfo{author}{\bibfnamefont{M.~I.} \bibnamefont{Katsnelson}},
  \bibnamefont{and} \bibinfo{author}{\bibfnamefont{K.~S.}
  \bibnamefont{Novoselov}}, \bibinfo{journal}{Nat. Mater.}
  \textbf{\bibinfo{volume}{6}}, \bibinfo{pages}{652} (\bibinfo{year}{2007}).

\bibitem[{\citenamefont{Novoselov}(2007)}]{Novoselov07}
\bibinfo{author}{\bibfnamefont{K.}~\bibnamefont{Novoselov}},
  \bibinfo{journal}{Nat. Mater.} \textbf{\bibinfo{volume}{6}},
  \bibinfo{pages}{720} (\bibinfo{year}{2007}).

\bibitem[{\citenamefont{Jeloaica and Sidis}(1999)}]{Jeloaica1999}
\bibinfo{author}{\bibfnamefont{L.}~\bibnamefont{Jeloaica}} \bibnamefont{and}
  \bibinfo{author}{\bibfnamefont{V.}~\bibnamefont{Sidis}},
  \bibinfo{journal}{Chem. Phys. Lett.} \textbf{\bibinfo{volume}{300}},
  \bibinfo{pages}{157} (\bibinfo{year}{1999}).

\bibitem[{\citenamefont{Sha and Jackson}(2002)}]{Sha&Jackson2002}
\bibinfo{author}{\bibfnamefont{X.}~\bibnamefont{Sha}} \bibnamefont{and}
  \bibinfo{author}{\bibfnamefont{B.}~\bibnamefont{Jackson}},
  \bibinfo{journal}{Surf. Sci.} \textbf{\bibinfo{volume}{496}},
  \bibinfo{pages}{318} (\bibinfo{year}{2002}).

\bibitem[{\citenamefont{Casolo et~al.}(2009)\citenamefont{Casolo, L{\o}vvik,
  Martinazzo, and Tantardini}}]{casolo09}
\bibinfo{author}{\bibfnamefont{S.}~\bibnamefont{Casolo}},
  \bibinfo{author}{\bibfnamefont{O.~M.} \bibnamefont{L{\o}vvik}},
  \bibinfo{author}{\bibfnamefont{R.}~\bibnamefont{Martinazzo}},
  \bibnamefont{and} \bibinfo{author}{\bibfnamefont{G.~F.}
  \bibnamefont{Tantardini}}, \bibinfo{journal}{J. Chem. Phys.}
  \textbf{\bibinfo{volume}{130}}, \bibinfo{pages}{054704}
  (\bibinfo{year}{2009}).

\bibitem[{\citenamefont{Hornek{\ae}r
  et~al.}(2006{\natexlab{a}})\citenamefont{Hornek{\ae}r,
  {\v{S}}ljivan{\v{c}}anin, Xu, Otero, Rauls, Stensgaard, L{\ae}gsgaard,
  Hammer, and Besenbacher}}]{Hornekaer2006}
\bibinfo{author}{\bibfnamefont{L.}~\bibnamefont{Hornek{\ae}r}},
  \bibinfo{author}{\bibfnamefont{{\v{Z}}.}~\bibnamefont{{\v{S}}ljivan{\v{c}}an%
in}}, \bibinfo{author}{\bibfnamefont{W.}~\bibnamefont{Xu}},
  \bibinfo{author}{\bibfnamefont{R.}~\bibnamefont{Otero}},
  \bibinfo{author}{\bibfnamefont{E.}~\bibnamefont{Rauls}},
  \bibinfo{author}{\bibfnamefont{I.}~\bibnamefont{Stensgaard}},
  \bibinfo{author}{\bibfnamefont{E.}~\bibnamefont{L{\ae}gsgaard}},
  \bibinfo{author}{\bibfnamefont{B.}~\bibnamefont{Hammer}}, \bibnamefont{and}
  \bibinfo{author}{\bibfnamefont{F.}~\bibnamefont{Besenbacher}},
  \bibinfo{journal}{Phys. Rev. Lett.} \textbf{\bibinfo{volume}{96}},
  \bibinfo{pages}{156104} (\bibinfo{year}{2006}{\natexlab{a}}).

\bibitem[{\citenamefont{Hornek{\ae}r
  et~al.}(2006{\natexlab{b}})\citenamefont{Hornek{\ae}r, Rauls, Xu,
  {\v{S}}ljivan{\v{c}}anin, Otero, Stensgaard, L{\ae}egsgaard, Hammer, and
  Besenbacher}}]{Hornekaer2006a}
\bibinfo{author}{\bibfnamefont{L.}~\bibnamefont{Hornek{\ae}r}},
  \bibinfo{author}{\bibfnamefont{E.}~\bibnamefont{Rauls}},
  \bibinfo{author}{\bibfnamefont{W.}~\bibnamefont{Xu}},
  \bibinfo{author}{\bibfnamefont{{\v{Z}}.}~\bibnamefont{{\v{S}}ljivan{\v{c}}an%
in}}, \bibinfo{author}{\bibfnamefont{R.}~\bibnamefont{Otero}},
  \bibinfo{author}{\bibfnamefont{I.}~\bibnamefont{Stensgaard}},
  \bibinfo{author}{\bibfnamefont{E.}~\bibnamefont{L{\ae}egsgaard}},
  \bibinfo{author}{\bibfnamefont{B.}~\bibnamefont{Hammer}}, \bibnamefont{and}
  \bibinfo{author}{\bibfnamefont{F.}~\bibnamefont{Besenbacher}},
  \bibinfo{journal}{Phys. Rev. Lett.} \textbf{\bibinfo{volume}{97}},
  \bibinfo{pages}{186102} (\bibinfo{year}{2006}{\natexlab{b}}).

\bibitem[{\citenamefont{Boukhvalov et~al.}(2008)\citenamefont{Boukhvalov,
  Katsnelson, and Lichtenstein}}]{Boukhvalov2008}
\bibinfo{author}{\bibfnamefont{D.~W.} \bibnamefont{Boukhvalov}},
  \bibinfo{author}{\bibfnamefont{M.~I.} \bibnamefont{Katsnelson}},
  \bibnamefont{and} \bibinfo{author}{\bibfnamefont{A.~I.}
  \bibnamefont{Lichtenstein}}, \bibinfo{journal}{Phys. Rev. B}
  \textbf{\bibinfo{volume}{77}}, \bibinfo{pages}{035427}
  (\bibinfo{year}{2008}).

\bibitem[{\citenamefont{Gould and Salpeter}(1963)}]{Gould63}
\bibinfo{author}{\bibfnamefont{R.~J.} \bibnamefont{Gould}} \bibnamefont{and}
  \bibinfo{author}{\bibfnamefont{E.~E.} \bibnamefont{Salpeter}},
  \bibinfo{journal}{Astrophys. J.} \textbf{\bibinfo{volume}{138}}
  (\bibinfo{year}{1963}).

\bibitem[{\citenamefont{Hollenbach and Salpeter}(1970)}]{Hollenbach70}
\bibinfo{author}{\bibfnamefont{D.}~\bibnamefont{Hollenbach}} \bibnamefont{and}
  \bibinfo{author}{\bibfnamefont{E.~E.} \bibnamefont{Salpeter}},
  \bibinfo{journal}{J. Chem. Phys.} \textbf{\bibinfo{volume}{53}},
  \bibinfo{pages}{79} (\bibinfo{year}{1970}).

\bibitem[{\citenamefont{Kresse and Hafner}(1994)}]{VASP1}
\bibinfo{author}{\bibfnamefont{G.}~\bibnamefont{Kresse}} \bibnamefont{and}
  \bibinfo{author}{\bibfnamefont{J.}~\bibnamefont{Hafner}},
  \bibinfo{journal}{Phys. Rev. B} \textbf{\bibinfo{volume}{49}},
  \bibinfo{pages}{14251} (\bibinfo{year}{1994}).

\bibitem[{\citenamefont{Kresse and Hafner}(1993)}]{VASP2}
\bibinfo{author}{\bibfnamefont{G.}~\bibnamefont{Kresse}} \bibnamefont{and}
  \bibinfo{author}{\bibfnamefont{J.}~\bibnamefont{Hafner}},
  \bibinfo{journal}{Phys. Rev. B} \textbf{\bibinfo{volume}{47}},
  \bibinfo{pages}{558} (\bibinfo{year}{1993}).

\bibitem[{\citenamefont{Bl{\"o}chl}(1994)}]{PAW1}
\bibinfo{author}{\bibfnamefont{P.~E.} \bibnamefont{Bl{\"o}chl}},
  \bibinfo{journal}{Phys. Rev. B} \textbf{\bibinfo{volume}{50}},
  \bibinfo{pages}{17953} (\bibinfo{year}{1994}).

\bibitem[{\citenamefont{Slonczewski and Weiss}(1958)}]{Weiss58}
\bibinfo{author}{\bibfnamefont{J.~C.} \bibnamefont{Slonczewski}}
  \bibnamefont{and} \bibinfo{author}{\bibfnamefont{P.~R.} \bibnamefont{Weiss}},
  \bibinfo{journal}{Phys. Rev.} \textbf{\bibinfo{volume}{109}},
  \bibinfo{pages}{272} (\bibinfo{year}{1958}).

\bibitem[{\citenamefont{Inui et~al.}(1994)\citenamefont{Inui, Trugman, and
  Abrahams}}]{Inui1994}
\bibinfo{author}{\bibfnamefont{M.}~\bibnamefont{Inui}},
  \bibinfo{author}{\bibfnamefont{S.~A.} \bibnamefont{Trugman}},
  \bibnamefont{and} \bibinfo{author}{\bibfnamefont{E.}~\bibnamefont{Abrahams}},
  \bibinfo{journal}{Phys. Rev. B} \textbf{\bibinfo{volume}{49}},
  \bibinfo{pages}{3190} (\bibinfo{year}{1994}).

\bibitem[{\citenamefont{Lieb}(1989)}]{lieb}
\bibinfo{author}{\bibfnamefont{E.~H.} \bibnamefont{Lieb}},
  \bibinfo{journal}{Phys. Rev. Lett.} \textbf{\bibinfo{volume}{62}},
  \bibinfo{pages}{1201} (\bibinfo{year}{1989}).

\bibitem[{\citenamefont{Wehling et~al.}(2009)\citenamefont{Wehling, Katsnelson,
  and Lichtenstein}}]{Wehling08}
\bibinfo{author}{\bibfnamefont{T.~O.} \bibnamefont{Wehling}},
  \bibinfo{author}{\bibfnamefont{M.~I.} \bibnamefont{Katsnelson}},
  \bibnamefont{and} \bibinfo{author}{\bibfnamefont{A.~I.}
  \bibnamefont{Lichtenstein}}, \bibinfo{journal}{Phys. Rev. B}
  \textbf{\bibinfo{volume}{80}}, \bibinfo{pages}{085428}
  (\bibinfo{year}{2009}).

\bibitem[{\citenamefont{Anderson}(1961)}]{NewnsAnderson2}
\bibinfo{author}{\bibfnamefont{P.~W.} \bibnamefont{Anderson}},
  \bibinfo{journal}{Phys. Rev.} \textbf{\bibinfo{volume}{124}},
  \bibinfo{pages}{41} (\bibinfo{year}{1961}).

\bibitem[{\citenamefont{Newns}(1969)}]{NewnsAnderson1}
\bibinfo{author}{\bibfnamefont{D.~M.} \bibnamefont{Newns}},
  \bibinfo{journal}{Phys. Rev.} \textbf{\bibinfo{volume}{178}},
  \bibinfo{pages}{1123} (\bibinfo{year}{1969}).

\bibitem[{\citenamefont{Lindenblatt and Pehlke}(2006)}]{Lindenblatt2006}
\bibinfo{author}{\bibfnamefont{M.}~\bibnamefont{Lindenblatt}} \bibnamefont{and}
  \bibinfo{author}{\bibfnamefont{E.}~\bibnamefont{Pehlke}},
  \bibinfo{journal}{Phys. Rev. Lett.} \textbf{\bibinfo{volume}{97}},
  \bibinfo{pages}{216101} (\bibinfo{year}{2006}).

\bibitem[{\citenamefont{M.S. et~al.}(2008)\citenamefont{M.S., Mizielinski,
  Bird, Persson, and Holloway}}]{Mizielinski2008}
\bibinfo{author}{\bibnamefont{M.S.}},
  \bibinfo{author}{\bibnamefont{Mizielinski}},
  \bibinfo{author}{\bibfnamefont{D.}~\bibnamefont{Bird}},
  \bibinfo{author}{\bibfnamefont{M.}~\bibnamefont{Persson}}, \bibnamefont{and}
  \bibinfo{author}{\bibfnamefont{S.}~\bibnamefont{Holloway}},
  \bibinfo{journal}{Surf. Sci.} \textbf{\bibinfo{volume}{602}},
  \bibinfo{pages}{2617} (\bibinfo{year}{2008}).

\bibitem[{\citenamefont{Cohen et~al.}(2008{\natexlab{a}})\citenamefont{Cohen,
  {Mori-Sanchez}, and Yang}}]{CohenScience}
\bibinfo{author}{\bibfnamefont{A.~J.} \bibnamefont{Cohen}},
  \bibinfo{author}{\bibfnamefont{P.}~\bibnamefont{{Mori-Sanchez}}},
  \bibnamefont{and} \bibinfo{author}{\bibfnamefont{W.}~\bibnamefont{Yang}},
  \bibinfo{journal}{Science} \textbf{\bibinfo{volume}{321}},
  \bibinfo{pages}{792} (\bibinfo{year}{2008}{\natexlab{a}}).

\bibitem[{\citenamefont{Gritsenko et~al.}({2000})\citenamefont{Gritsenko,
  T.Schipper, and Baerends}}]{Gritsenko2000}
\bibinfo{author}{\bibfnamefont{O.}~\bibnamefont{Gritsenko}},
  \bibinfo{author}{\bibfnamefont{B.~E. P.~R.} \bibnamefont{T.Schipper}},
  \bibnamefont{and} \bibinfo{author}{\bibfnamefont{E.~J.}
  \bibnamefont{Baerends}}, \bibinfo{journal}{J. Phys. Chem. A}
  \textbf{\bibinfo{volume}{{104}}}, \bibinfo{pages}{8558}
  (\bibinfo{year}{{2000}}).

\bibitem[{\citenamefont{Zhang and Yang}(1998)}]{ZhangYang98}
\bibinfo{author}{\bibfnamefont{Y.}~\bibnamefont{Zhang}} \bibnamefont{and}
  \bibinfo{author}{\bibfnamefont{W.}~\bibnamefont{Yang}}, \bibinfo{journal}{J.
  Chem. Phys.} \textbf{\bibinfo{volume}{109}}, \bibinfo{pages}{2604}
  (\bibinfo{year}{1998}).

\bibitem[{\citenamefont{{Mori-Sanchez}
  et~al.}(2009)\citenamefont{{Mori-Sanchez}, Cohen, and Yang}}]{MoriSanchez09}
\bibinfo{author}{\bibfnamefont{P.}~\bibnamefont{{Mori-Sanchez}}},
  \bibinfo{author}{\bibfnamefont{A.}~\bibnamefont{Cohen}}, \bibnamefont{and}
  \bibinfo{author}{\bibfnamefont{W.}~\bibnamefont{Yang}},
  \bibinfo{journal}{Phys. Rev. Lett.} p. \bibinfo{pages}{066403}
  (\bibinfo{year}{2009}).

\bibitem[{\citenamefont{Burke et~al.}(1998)\citenamefont{Burke, Perdew, and
  Ernzerhof}}]{Burke98}
\bibinfo{author}{\bibfnamefont{K.}~\bibnamefont{Burke}},
  \bibinfo{author}{\bibfnamefont{J.~P.} \bibnamefont{Perdew}},
  \bibnamefont{and}
  \bibinfo{author}{\bibfnamefont{M.}~\bibnamefont{Ernzerhof}},
  \bibinfo{journal}{J. Chem. Phys.} \textbf{\bibinfo{volume}{109}},
  \bibinfo{pages}{3760} (\bibinfo{year}{1998}).

\bibitem[{\citenamefont{Perdew et~al.}(1995)\citenamefont{Perdew, Savin, and
  Burke}}]{Perdew95}
\bibinfo{author}{\bibfnamefont{J.~P.} \bibnamefont{Perdew}},
  \bibinfo{author}{\bibfnamefont{A.}~\bibnamefont{Savin}}, \bibnamefont{and}
  \bibinfo{author}{\bibfnamefont{K.}~\bibnamefont{Burke}},
  \bibinfo{journal}{Phys. Rev. A} \textbf{\bibinfo{volume}{51}},
  \bibinfo{pages}{4531} (\bibinfo{year}{1995}).

\bibitem[{\citenamefont{Perdew et~al.}(1982)\citenamefont{Perdew, Parr, Levy,
  and {Balduz Jr.}}}]{Perdew1982}
\bibinfo{author}{\bibfnamefont{G.~B.} \bibnamefont{Perdew}},
  \bibinfo{author}{\bibfnamefont{R.~G.} \bibnamefont{Parr}},
  \bibinfo{author}{\bibfnamefont{M.}~\bibnamefont{Levy}}, \bibnamefont{and}
  \bibinfo{author}{\bibfnamefont{J.~L.} \bibnamefont{{Balduz Jr.}}},
  \bibinfo{journal}{Phys. Rev. Lett.} \textbf{\bibinfo{volume}{49}},
  \bibinfo{pages}{1691} (\bibinfo{year}{1982}).

\bibitem[{\citenamefont{Perdew et~al.}(2007)\citenamefont{Perdew, Ruzsinsky,
  Csonka, Vydrov, and Scuseria}}]{Perdew07}
\bibinfo{author}{\bibfnamefont{J.~P.} \bibnamefont{Perdew}},
  \bibinfo{author}{\bibfnamefont{A.}~\bibnamefont{Ruzsinsky}},
  \bibinfo{author}{\bibfnamefont{G.~I.} \bibnamefont{Csonka}},
  \bibinfo{author}{\bibfnamefont{O.~A.} \bibnamefont{Vydrov}},
  \bibnamefont{and} \bibinfo{author}{\bibfnamefont{G.~E.}
  \bibnamefont{Scuseria}}, \bibinfo{journal}{Phys. Rev. A}
  \textbf{\bibinfo{volume}{76}}, \bibinfo{pages}{040501}
  (\bibinfo{year}{2007}).

\bibitem[{\citenamefont{{Mori-Sanchez}
  et~al.}(2008)\citenamefont{{Mori-Sanchez}, Cohen, and Yang}}]{MoriSanchez08}
\bibinfo{author}{\bibfnamefont{P.}~\bibnamefont{{Mori-Sanchez}}},
  \bibinfo{author}{\bibfnamefont{A.}~\bibnamefont{Cohen}}, \bibnamefont{and}
  \bibinfo{author}{\bibfnamefont{W.}~\bibnamefont{Yang}},
  \bibinfo{journal}{Phys. Rev. Lett.} \textbf{\bibinfo{volume}{100}},
  \bibinfo{pages}{146401} (\bibinfo{year}{2008}).

\bibitem[{\citenamefont{Ruzsinszky et~al.}(2006)\citenamefont{Ruzsinszky,
  Perdew, Csonka, Vydrov, and Scuseria}}]{Ruzsinszky06}
\bibinfo{author}{\bibfnamefont{A.}~\bibnamefont{Ruzsinszky}},
  \bibinfo{author}{\bibfnamefont{J.~P.} \bibnamefont{Perdew}},
  \bibinfo{author}{\bibfnamefont{G.~I.} \bibnamefont{Csonka}},
  \bibinfo{author}{\bibfnamefont{O.~A.} \bibnamefont{Vydrov}},
  \bibnamefont{and} \bibinfo{author}{\bibfnamefont{G.~E.}
  \bibnamefont{Scuseria}}, \bibinfo{journal}{J. Chem. Phys.}
  \textbf{\bibinfo{volume}{125}}, \bibinfo{pages}{194112}
  (\bibinfo{year}{2006}).

\bibitem[{\citenamefont{Marsman et~al.}(2008)\citenamefont{Marsman, Paier,
  Stroppa, and Kresse}}]{Marsman08}
\bibinfo{author}{\bibfnamefont{M.}~\bibnamefont{Marsman}},
  \bibinfo{author}{\bibfnamefont{J.}~\bibnamefont{Paier}},
  \bibinfo{author}{\bibfnamefont{A.}~\bibnamefont{Stroppa}}, \bibnamefont{and}
  \bibinfo{author}{\bibfnamefont{G.}~\bibnamefont{Kresse}},
  \bibinfo{journal}{J. Phys.: Condens. Matter} \textbf{\bibinfo{volume}{20}},
  \bibinfo{pages}{064201} (\bibinfo{year}{2008}).

\bibitem[{\citenamefont{Adamo and Barone}(1999)}]{PBE0}
\bibinfo{author}{\bibfnamefont{C.}~\bibnamefont{Adamo}} \bibnamefont{and}
  \bibinfo{author}{\bibfnamefont{V.}~\bibnamefont{Barone}},
  \bibinfo{journal}{J. Chem. Phys.} \textbf{\bibinfo{volume}{110}},
  \bibinfo{pages}{6158} (\bibinfo{year}{1999}).

\bibitem[{\citenamefont{Cohen et~al.}(2008{\natexlab{b}})\citenamefont{Cohen,
  Mori-Sanchez, and Yang}}]{CohenBandGap}
\bibinfo{author}{\bibfnamefont{A.}~\bibnamefont{Cohen}},
  \bibinfo{author}{\bibfnamefont{P.}~\bibnamefont{Mori-Sanchez}},
  \bibnamefont{and} \bibinfo{author}{\bibfnamefont{W.}~\bibnamefont{Yang}},
  \bibinfo{journal}{Phys. Rev. B} \textbf{\bibinfo{volume}{77}},
  \bibinfo{pages}{115123} (\bibinfo{year}{2008}{\natexlab{b}}).

\bibitem[{\citenamefont{Dudarev et~al.}(1998)\citenamefont{Dudarev, Botton,
  Savrasov, Humphreys, and Sutton}}]{LSDAU}
\bibinfo{author}{\bibfnamefont{S.~L.} \bibnamefont{Dudarev}},
  \bibinfo{author}{\bibfnamefont{G.~A.} \bibnamefont{Botton}},
  \bibinfo{author}{\bibfnamefont{S.~Y.} \bibnamefont{Savrasov}},
  \bibinfo{author}{\bibfnamefont{C.~J.} \bibnamefont{Humphreys}},
  \bibnamefont{and} \bibinfo{author}{\bibfnamefont{A.~P.}
  \bibnamefont{Sutton}}, \bibinfo{journal}{Phys. Rev. B}
  \textbf{\bibinfo{volume}{57}}, \bibinfo{pages}{1505} (\bibinfo{year}{1998}).

\bibitem[{\citenamefont{Cococcioni and de~Gironcoli}(2005)}]{Cococcioni05}
\bibinfo{author}{\bibfnamefont{M.}~\bibnamefont{Cococcioni}} \bibnamefont{and}
  \bibinfo{author}{\bibfnamefont{S.}~\bibnamefont{de~Gironcoli}},
  \bibinfo{journal}{Phys. Rev. B} \textbf{\bibinfo{volume}{71}},
  \bibinfo{pages}{035105} (\bibinfo{year}{2005}).

\bibitem[{\citenamefont{Pedersen}(2004)}]{Pedersen04}
\bibinfo{author}{\bibfnamefont{T.~G.} \bibnamefont{Pedersen}},
  \bibinfo{journal}{Phys. Rev. B} \textbf{\bibinfo{volume}{69}},
  \bibinfo{pages}{075207} (\bibinfo{year}{2004}).

\bibitem[{\citenamefont{Pedersen et~al.}(2008)\citenamefont{Pedersen, Flindt,
  Pedersen, Mortensen, Jauho, and Pedersen}}]{Pedersen08}
\bibinfo{author}{\bibfnamefont{T.}~\bibnamefont{Pedersen}},
  \bibinfo{author}{\bibfnamefont{C.}~\bibnamefont{Flindt}},
  \bibinfo{author}{\bibfnamefont{J.}~\bibnamefont{Pedersen}},
  \bibinfo{author}{\bibfnamefont{N.}~\bibnamefont{Mortensen}},
  \bibinfo{author}{\bibfnamefont{A.~P.} \bibnamefont{Jauho}}, \bibnamefont{and}
  \bibinfo{author}{\bibfnamefont{K.}~\bibnamefont{Pedersen}},
  \bibinfo{journal}{Phys. Rev. Lett.} \textbf{\bibinfo{volume}{100}},
  \bibinfo{pages}{136804} (\bibinfo{year}{2008}).

\end{thebibliography}

\end{document}